\documentclass[11pt]{article}

\usepackage{amsmath,amssymb,bbm,amsthm}
\usepackage{fullpage}
\usepackage{thm-restate,color,xcolor,xspace,graphicx}
\usepackage{hyperref,cleveref}
\usepackage{algorithm,algorithmicx}
\usepackage[noend]{algpseudocode}
\usepackage{balance}

\AtBeginDocument{%
  \providecommand\BibTeX{{%
    \normalfont B\kern-0.5em{\scshape i\kern-0.25em b}\kern-0.8em\TeX}}}

\newcommand{\f}{\frac}
\newcommand{\cd}{\cdot}
\newcommand{\bn}{\binom}

\newcommand{\cds}{\cdots}
\newcommand{\lds}{\ldots}

\newcommand{\sm}{\setminus}
\newcommand{\s}{\subseteq}

\newcommand{\BE}{\begin{enumerate}}
\newcommand{\EE}{\end{enumerate}}
\newcommand{\im}{\item}
\newcommand{\BI}{\begin{itemize}}
\newcommand{\EI}{\end{itemize}}
\def\BAL#1\EAL{\begin{align*}#1\end{align*}}
\def\BALN#1\EALN{\begin{align}#1\end{align}}
\def\BG#1\EG{\begin{gather}#1\end{gather}}

\newcommand{\logn}{\log n}

\newcommand{\inv}{^{-1}}

\newcommand{\R}{\mathbb R}

\newcommand{\e}{\epsilon}
\newcommand{\de}{\delta}
\newcommand{\De}{\Delta}
\newcommand{\la}{\lambda}

\newcommand{\pt}{\partial}

\newcommand{\Om}{\Omega}

\newcommand{\Th}{\Theta}

\newcommand{\lf}{\lfloor}
\newcommand{\rf}{\rfloor}

\newcommand{\E}{\mathbb E}

\newcommand{\poly}{\textup{poly}}
\newcommand{\polylog}{\textup{polylog}}
\newcommand{\pl}{\textup{polylog}}

\newcommand{\lp}{\left(}
\newcommand{\rp}{\right)}

\newcommand{\lmt}{\left[\begin{matrix}}
\newcommand{\rmt}{\end{matrix}\right]}

\newtheorem{theorem}{Theorem}[section]

\newtheorem{lemma}[theorem]{Lemma}
\newtheorem{definition}[theorem]{Definition}
\newtheorem{corollary}[theorem]{Corollary}
\newtheorem{observation}[theorem]{Observation}
\newtheorem{claim}[theorem]{Claim}
\newtheorem{subclaim}[theorem]{Subclaim}
\newtheorem{fact}[theorem]{Fact}
\newtheorem{assumption}[theorem]{Assumption}

\newcommand{\BT}{\begin{theorem}}
\newcommand{\ET}{\end{theorem}}
\newcommand{\BL}{\begin{lemma}}
\newcommand{\EL}{\end{lemma}}
\newcommand{\BD}{\begin{definition}}
\newcommand{\ED}{\end{definition}}
\newcommand{\BC}{\begin{corollary}}
\newcommand{\EC}{\end{corollary}}
\newcommand{\BO}{\begin{observation}}
\newcommand{\EO}{\end{observation}}
\newcommand{\BCL}{\begin{claim}}
\newcommand{\ECL}{\end{claim}}
\newcommand{\BSCL}{\begin{subclaim}}
\newcommand{\ESCL}{\end{subclaim}}
\newcommand{\BF}{\begin{fact}}
\newcommand{\EF}{\end{fact}}
\newcommand{\BA}{\begin{assumption}}
\newcommand{\EA}{\end{assumption}}
\newcommand{\BP}{\begin{proof}}
\newcommand{\EP}{\end{proof}}
\newcommand{\BSP}{\begin{subproof}}
\newcommand{\ESP}{\end{subproof}}
\newcommand{\BPS}{\begin{proof}[Proof (Sketch)]}
\newcommand{\EPS}{\end{proof}}
\Crefname{observation}{Observation}{Observations}
\Crefname{claim}{Claim}{Claims}
\Crefname{subclaim}{Subclaim}{Subclaims}
\Crefname{fact}{Fact}{Facts}
\Crefname{assumption}{Assumption}{Assumptions}

\newenvironment{subproof}[1][\proofname]{%
  \begin{proof}[#1]%
}{%
  \end{proof}%
}

\newcommand{\eat}[1]{}

\newcommand{\tO}{\tilde{O}}

\newcommand{\thml}[1]{\label{thm:#1}}
\newcommand{\thm}[1]{\Cref{thm:#1}}
\newcommand{\leml}[1]{\label{lem:#1}}
\newcommand{\lem}[1]{\Cref{lem:#1}}

\newcommand{\corl}[1]{\label{cor:#1}}
\newcommand{\cor}[1]{\Cref{cor:#1}}

\newcommand{\linel}[1]{\label{line:#1}}
\renewcommand{\line}[1]{line~\ref{line:#1}}
\newcommand{\secl}[1]{\label{sec:#1}}
\renewcommand{\sec}[1]{\Cref{sec:#1}}
\makeatletter
\newcounter{algocounter}
\@ifpackageloaded{hyperref}%
  {\newcommand{\mylabel}[2]% #1=name, #2 = contents
    {\refstepcounter{algocounter}\protected@write\@auxout{}{\string\newlabel{#1}{{\textcolor{black}{\textup{#2}}}{\thepage}%
      {\@currentlabelname}{\@currentHref}{}}}}}%
\makeatother

\newcommand{\mincut}{\textsf{\textup{mincut}}}

\newcommand{\sma}{{\textup{small}}}
\newcommand{\lar}{{\textup{large}}}

\newcommand{\ssc}{{\sc SSMC}\xspace}
\newcommand{\apc}{{\sc APMC}\xspace}
\newcommand{\ct}{{\sc CT}\xspace}

\begin{document}

\title{Approximate Gomory--Hu Tree Is Faster Than $n-1$ Max-Flows}
\author{
Jason Li\\Carnegie Mellon University\\Email: {\tt jmli@cs.cmu.edu}
\and 
Debmalya Panigrahi\\Duke University\\Email: {\tt debmalya@cs.duke.edu}
}
\date{}
\maketitle

\begin{abstract}
    The Gomory-Hu tree or cut tree (Gomory and Hu, 1961) is a classic data structure for reporting $(s,t)$ mincuts (and by duality, the values of $(s,t)$ maxflows) for all pairs of vertices $s$ and $t$ in an undirected graph. Gomory and Hu showed that it can be computed using $n-1$ exact maxflow computations. Surprisingly, this remains the best algorithm for Gomory-Hu trees more than 50 years later, {\em even for approximate mincuts}. In this paper, we break this longstanding barrier and give an algorithm for computing a $(1+\e)$-approximate Gomory-Hu tree using $\polylog(n)$ maxflow computations. Specifically, we obtain the runtime bounds we describe below.
    
    We obtain a randomized (Monte Carlo) algorithm for undirected, weighted graphs that runs in $\tO(m + n^{3/2})$ time and returns a $(1+\e)$-approximate Gomory-Hu tree algorithm w.h.p. Previously, the best running time known was $\tO(n^{5/2})$, which is obtained by running Gomory and Hu's original algorithm on a cut sparsifier of the graph.
    
    Next, we obtain a randomized (Monte Carlo) algorithm for undirected, unweighted graphs that runs in $m^{4/3+o(1)}$ time and returns a $(1+\e)$-approximate Gomory-Hu tree algorithm w.h.p. This improves on our first result for sparse graphs, namely $m = o(n^{9/8})$. Previously, the best running time known for unweighted graphs was $\tO(mn)$ for an exact Gomory-Hu tree (Bhalgat {\em et al.}, STOC 2007); no better result was known if approximations are allowed.
    
    As a consequence of our Gomory-Hu tree algorithms, we also solve the $(1+\e)$-approximate all pairs mincut (\apc) and single source mincut (\ssc) problems in the same time bounds. (These problems are simpler in that the goal is to only return the $(s,t)$ mincut values, and not the mincuts.) This improves on the recent algorithm for these problems in $\tO(n^2)$ time due to Abboud~{\em et al.} (FOCS 2020).
\end{abstract}

\section{Introduction}
\label{sec:introduction}
The algorithmic study of cuts and flows is one of the pillars of combinatorial optimization. The foundations of this field were established 
in the celebrated work of Ford and Fulkerson in the mid-50s~\cite{FordF56}. They studied the $(s,t)$ edge connectivity problem, namely finding a set of edges of minimum weight whose removal disconnects two vertices $s$ from $t$ in a graph (such a set of edges is called an $(s,t)$-mincut). They showed that the weight of an $(s,t)$-mincut equals the maximum flow between $s$ and $t$ in the graph, a duality that has underpinned much of the success in this field. Soon after their work, in a remarkable result, Gomory and Hu~\cite{GomoryH61} showed that by using just $n-1$ maxflows, they could construct a tree $T$ on the vertices of an undirected graph $G$ such that for every pair of vertices $s$ and $t$, the $(s,t)$ edge connectivity in $T$ was equal to that in $G$. In other words, the $\binom n 2$ pairs of vertices had at most $n-1$ different edge connectivities and they could be obtained using just $n-1$ maxflow calls. Moreover, for all vertex pairs $s$ and $t$, the bipartition of vertices in the $(s,t)$-mincut in tree $T$ (note that this is just the bipartition created by removing the minimum weight edge on the unique $s-t$ path in $T$) was also an $(s,t)$-mincut in graph $G$. This data structure, called a {\em cut tree} or more appropriately a {\em Gomory-Hu tree} (abbreviated GH-tree) after its creators, has become a standard feature in algorithms textbooks, courses, and research since their work.

But, rather surprisingly, in spite of the remarkable successes in this field as a whole, the best algorithm for constructing a GH-tree remains the one given by Gomory and Hu almost six decades after their work. There have been alternatives suggested along the way, although none of them unconditionally improves on the original construction. Gusfield~\cite{Gusfield90} gave an algorithm that also uses $n-1$ maxflows, but on the original graph itself (the GH algorithm runs maxflows on contracted graphs as we will see later) to improve the performance of the algorithm in practice. Bhalgat~{\em et al.}~\cite{BhalgatHKP08} (see also \cite{HariharanKP07}) obtained an $\tO(mn)$ algorithm for this problem, but only for unweighted graphs. (Note that using the state of the art maxflow algorithms~\cite{liu2020faster}, the GH algorithm has a running time of $m^{4/3+o(1)}n$ for unweighted graphs, which is slower.) Karger and Levine~\cite{KargerL15} matched this running time using a randomized maxflow subroutine, also for unweighted graphs. Recently, Abboud~{\em et al.}~\cite{AbboudKT20a} improved this bound for {\em sparse} unweighted graphs to $\tO(m^{3/2}n^{1/6})$, thereby demonstrating that the $\tO(mn)$ is not tight, at least in certain edge density regimes. Further improvements have been obtained in special cases: in particular, near-linear time algorithms are known for planar graphs~\cite{BorradaileSW15} and surface-embedded graphs~\cite{BorradaileENW16}. Experimental studies of GH tree algorithms have also been performed~\cite{GoldbergT01}. The reader is referred to a survey article on this topic for more background~\cite{Panigrahi16}.

In spite of all the works described above, the status of the GH tree problem for general weighted graphs has remained unchanged for the last six decades. Namely, we know that a GH tree can be constructed using $n-1$ maxflows, but can we do better? In fact, surprisingly, a faster GH tree algorithm is not known {\em even if one allowed approximations}, i.e., if the $(s,t)$-mincuts in the GH tree and those in the original graph could differ by a multiplicative factor. At first glance, this would appear surprising, since $\tO(m)$-time algorithms for  $(1+\e)$-approximation of maxflows are known. (In contrast, obtaining an exact maxflow algorithm that runs in near-linear time remains one of the major open challenges in graph algorithms.) But, the difficulty in using these faster approximate maxflow algorithms in the GH tree problem is that the GH algorithm (and also Gusfield's algorithm) use recursive calls in a manner that approximation errors can build up across the different recursive layers of the algorithm. Approximation, however, does present some advantage, in that one can use standard graph sparsification techniques to reduce the number of edges to $\tO(n)$ (see, e.g., \cite{BenczurK15,FungHHP19}) and then apply the GH algorithm (with exact maxflow) on this sparse graph. This reduces the running time to $n-1$ invocations of maxflow on $\tO(n)$-edge graphs, which has a total running time of $\tO(n^{5/2})$ using the current state of the art maxflow algorithm of Lee and Sidford~\cite{LeeSflow}. But, fundamentally, even allowing approximations, we do not have a GH tree algorithm that beats the $O(n)$ maxflows benchmark set by the original GH algorithm.

But, there has been some exciting progress of late in this line of research. Very recently, in a beautiful paper, Abboud~{\em et al.}~\cite{AbboudKT20b} showed that the problem of finding all pairs edge connectivities (that a GH tree obtains) can be reduced to $\polylog(n)$ instances of the single source mincut problem (we call this the \ssc problem). Given a fixed source vertex $s$, the latter problem asks for the $(s,t)$ edge connectivity of $s$ with every other vertex $t$. Their reduction is also robust to approximations because, crucially, the recursive depth of the reduction is only $\polylog(n)$ (as against the recursive depth of GH and Gusfield's algorithms, which can be $\Omega(n)$). So, in essence, they reduced the recursive depth of the algorithm in exchange for using a more powerful primitive, namely edge connectivity for $n-1$ pairs of vertices (one of the pair is common) rather than for just a single pair. The algorithm that they used to solve the single source edge connectivity problem is the obvious one: run $(s,t)$ maxflow for every vertex $t$. Naturally, this does not improve the running time for exact all pairs edge connectivity, since we are still running $n-1$ maxflows. But, importantly, if approximations are allowed, we can now use the $\tO(m)$-time approximate maxflow algorithm rather than the exact one. Coupled with sparsification, this yields a running time bound of $\tO(n^2)$ improving on the previous bound of $\tO(n^{5/2})$. 

However, while this improves the time complexity of approximate all pairs edge connectivity, the reduction framework of \cite{AbboudKT20b} does not support the construction of an approximate GH tree. Namely, they give a data structure (called a {\em flow tree}) that returns the (approximate) edge connectivity of a vertex pair when queried, but does not return a mincut for that pair. Nevertheless, this result creates a range of possibilities, now that we have a technique for designing computation trees for all pairs edge connectivity that have small recursive depth. In this paper, we give the first approximation algorithm (our approximation factor is $1+\e$ for any $\e>0$) for GH tree that beats the running time of $n-1$ maxflow calls. Namely, we show that a $(1+\e)$-approximate GH-tree can be constructed using polylog number of calls to an exact maxflow subroutine, plus $\tO(m)$ time outside these maxflow calls.

\subsection{Our Results}

%\alert{DP: Should we formally define the problem statements here?} \textcolor{blue}{JL: not sure, since the notation for approx GH tree is a bit cumbersome. Maybe just say "see Definition XX for the formal def for approx GH tree"?}

To state our main result, we first formally define an approximate GH tree. 

\BD[Approximate Gomory-Hu tree]
Given a graph $G=(V,E)$, a $(1+\e)$-approximate Gomory-Hu tree is a weighted tree $T$ on $V$ such that
 \BI
 \im For all $s,t\in V$, consider the minimum-weight edge $(u,v)$ on the unique $s-t$ path in $T$. Let $U'$ be the vertices of the connected component of $T-(u,v)$ containing $s$.
Then, the set $U'\s V$ is a $(1+\e)$-approximate $(s,t)$-mincut, and its value is the weight of the $(u, v)$ edge in $T$.
 \EI
\ED

We now state our main theorem that obtains a $(1+\e)$-approximate GH tree for weighted graphs:
\begin{restatable}{theorem}{ApproxW}\thml{approx-w}
    Let $G$ be an undirected graph with non-negative edge weights. There is a randomized algorithm that w.h.p., outputs a $(1+\e)$-approximate Gomory-Hu tree and runs in $\tO(m)$ time plus calls to exact max-flow on instances with a total of $\tO(n\e\inv\log^2\De)$ vertices and $\tO(n\e\inv\log^2\De)$ edges, where $\De$ is the ratio of maximum to minimum edge weights. Assuming polynomially bounded edge weights and using the $\tO(m\sqrt{n})$ time max-flow algorithm of Lee and Sidford~\cite{LeeSflow}, the algorithm runs in $\tO(m + n^{3/2}\e^{-2})$ time.
\end{restatable}
For unweighted graphs, we obtain the following result, which gives a better running time for sparse graphs (if $m = o(n^{9/8})$):
\begin{restatable}{theorem}{ApproxU}\thml{approx-u}
Let $G$ be an unweighted, undirected graph. There is a randomized algorithm that w.h.p., outputs a $(1+\e)$-approximate Gomory-Hu tree and runs in $\tO(m)$ time plus calls to exact max-flow on unweighted instances with a total of $\tO(n\e\inv)$ vertices and $\tO(m\e\inv)$ edges. Using the $m^{4/3+o(1)}$-time max-flow algorithm for unweighted graphs of Liu and Sidford~\cite{liu2020faster}, the algorithm runs in $m^{4/3+o(1)}\e\inv$ time.
\end{restatable}

To the best of our knowledge, this is the first algorithm for (approximate) GH tree that goes beyond $n-1$ maxflow calls in general weighted graphs. Our reduction to exact maxflow instances is ``black box'', i.e., any maxflow algorithm can be used; as a consequence, if one were to assume that eventually maxflow would be solved in $\tO(m)$-time as is often conjectured, then these theorems would automatically yield an $\tO(m)$-time algorithm for a $(1+\e)$-approximate GH tree.

Given these results, one might be tempted to replace the exact maxflow calls in our algorithm by approximate maxflow subroutines. Indeed, if this were possible, the running time of the overall algorithm would be $\tO(m)$ without additional assumptions (i.e., without assuming a $\tO(m)$-time exact maxflow algorithm). Unfortunately, a key tool that we employ called the {\em isolating cuts lemma}, which was recently introduced by the authors for the deterministic mincut problem~\cite{LiP20}, requires the computation of exact maxflows; we are not aware of any approximation versions of this lemma. We leave the problem of obtaining a near-linear time approximate GH tree algorithm as an interesting open question (that is probably easier than an exact $\tO(m)$-time maxflow algorithm).

Abboud~{\em et al.}~\cite{AbboudKT20b} recently considered the \apc (also called {\em flow tree}) problem, which asks for the value of the $(s,t)$-mincut for all vertex pairs $s, t$ but not a mincut itself. 

\BD[All-pairs min-cut]
In the \emph{all-pairs min-cut} (\apc) problem, the input is an undirected graph $G=(V,E)$ and we need to output a data structure that allows us to query the value of the $(s,t)$-mincut for each pair $s, t\in V$. In the \emph{$(1+\e)$-approximate \apc} problem, the input is the same, and we need to output a $(1+\e)$-approximation to the value of the $(s,v)$-mincut for each $v\in V\sm \{s\}$.
\ED

Abboud~{\em et al.} gave a framework that reduces the \apc problem to $\polylog(n)$ calls to the single source mincut (\ssc) problem.

\BD[Single-source min-cut]
In the \emph{single-source min-cut} (\ssc) problem, the input is an undirected graph $G=(V,E)$ and a source vertex $s\in V$, and we need to output a $(s,v)$-mincut for each $v\in V\sm \{s\}$. In the \emph{$(1+\e)$-approximate \ssc} problem, the input is the same, and we need to output a $(1+\e)$-approximate $(s,v)$-mincut for each $v\in V\sm \{s\}$.
\ED

To solve the \ssc instances, Abboud {\em et al.} used $n-1$ maxflows. Our work shows that the \ssc problem can be approximately solved using $\polylog(n)$ maxflows calls, and that an approximate GH tree can be recovered in the process. Our main tool is the following subroutine that we call the {\em Cut Threshold} (\ct) problem, %\alert{better name?}, 
which may have further applications on its own:

\begin{restatable}[Cut Threshold algorithm]{theorem}{Thr}\thml{thr}
Let $G=(V,E)$ be a weighted, undirected graph, and let $s\in V$, and let $\la\ge0$ be a parameter (the ``cut threshold"). There is an algorithm that outputs w.h.p. all vertices $v\in V$ with $\mincut(s,v)\le\la$, and runs in $\tO(m)$ time plus $\pl(n)$ calls to max-flow instances on $O(n)$-vertex, $O(m)$-edge graphs.
\end{restatable}

We use this theorem to obtain an algorithm for approximately solving the \ssc problem that is faster than running approximate maxflows for all the $n-1$ vertices separately:

\begin{restatable}{theorem}{SSMC}\thml{ssmc}
Let $G$ be a weighted, undirected graph, and let $s\in V$. There is an algorithm that outputs, for each vertex $v\in V\sm\{s\}$, a $(1+\e)$-approximation of $\mincut(s,v)$, and runs in $\tO(m\log\De)$ time plus $\pl(n)\cdot \log\De$ calls to max-flow on $O(n)$-vertex, $O(m)$-edge graphs, where $\De$ is the ratio of maximum to minimum edge weights.
\end{restatable}

Finally, note that a (approximate) GH tree also solves the (approximate) \apc problem. But, we can also get an \apc algorithm by simply plugging in the \ssc algorithm in \thm{ssmc} to the reduction framework of Abboud {\em et al.} This improves the time complexity of the \apc problem from $\tO(mn)$ obtained by  Abboud {\em et al.}~\cite{AbboudKT20b} to $\tO(m + n^{3/2})$.

%\textcolor{blue}{JL: should mention more work to be done to get GH tree}

\subsection{Our Techniques}

To sketch our main ideas, let us first think of the \ct problem (\thm{thr}). Note that this theorem is already sufficient to obtain the improved the running times for the \ssc and \apc problems, although obtaining a $(1+\e)$-approximate GH tree needs additional ideas. To solve the \ct problem, our main tool is the {\em isolating cuts lemma}, introduced by the authors recently for solving the deterministic mincut problem~\cite{LiP20}. We first describe this tool.\footnote{
We remark that the \emph{minimal} condition was not present in \cite{LiP20}, but the algorithm to find minimum isolating cuts from \cite{LiP20} can be trivially modified to output the minimal $(v,R\sm v)$-mincuts, so we omit the algorithm and direct interested readers to \cite{LiP20}.}
\BD[Minimum isolating cuts]
Consider a weighted, undirected graph $G=(V,E)$ and a subset $R\s V$ ($|R|\ge2$). The \emph{minimum isolating cuts} for $R$ is a collection of sets $\{S_v:v\in R\}$ such that for each vertex $v\in R$, the set $S_v$ is the side containing $v$ of the minimal $(v,R\sm v)$-mincut, i.e., for any set $S$ satisfying $v\in S$ and $S\cap(R\sm v)=\emptyset$, we have $w(\pt S)\le w(\pt S_v)$, and moreover, if $w(\pt S)=w(\pt S_v)$ then $S_v\s S$.
\ED
%
%\alert{DP: In the above definition, should we also mention that sets $S_v$ must be disjoint? At the moment, this is in the next lemma. Would putting it in the definition might make life a bit easier in terms of just having to refer to the definition when invoking this property?}
%
\BL [Isolating Cuts Lemma~\cite{LiP20})]
Fix a subset $R\s V$ ($|R|\ge2$). There is an algorithm that computes the minimum isolating cuts $\{S_v:v\in R\}$ for $R$ using $O(\log|R|)$ calls to $s$--$t$ max-flow on weighted graphs of $O(n)$ vertices and $O(m)$ edges, and takes $\tO(m)$ deterministic time outside of the max-flow calls. If the original graph $G$ is unweighted, then the inputs to the max-flow calls are also unweighted. Moreover, the sets $\{S_v:v\in R\}$ are disjoint.
\EL
The crucial aspect of the isolating cuts lemma is that the number of maxflow calls is $O(\log n)$ {\em irrespective of the size of $R$}. For the \ct problem, define $Z= V\setminus \{s\}$; our goal is to invoke the isolating cuts lemma $\polylog(n)$ times and identify all vertices $v\in Z$ with $\mincut(s, v)\le \lambda$ w.h.p. In fact, we will only describe an algorithm that identifies, in expectation, an $\Om(1/\logn)$ fraction of vertices in $Z$ satisfying this condition; removing these vertices from $Z$ and repeating $O(\log n)$ times identifies all such vertices in $Z$ w.h.p. Fix a vertex $v\in Z$ with $\mincut(s,v)\le\la$ and consider the minimal $(v,s)$-mincut, i.e., the $(v,s)$-mincut whose side $S_v\s V$ containing $v$ is inclusion-wise minimal. Let $n_v$ be the number of vertices in $Z$ on the side $S_v$. Now, suppose we sample a set of vertices from $Z$ at rate $1/n_v$ and define this sample as $R$. Then, we invoke the isolating cuts lemma with the set $R$, after adding $s$ to this set. Next, if the isolating cuts lemma returns cuts of value $\le \lambda$, we mark the vertices in $Z$ separated by those cuts from $s$ as having $\mincut(s, z)\le \lambda$ and remove them from $Z$. Clearly, every marked vertex $z$ indeed has $\min(s, z)\le \lambda$. But, how many vertices do we end up marking? Let us focus on the side $S_v$ of the minimal $(s,z)$-mincut. With constant probability, exactly one vertex from $S_v$ is sampled in $R$, and with probability $\Omega(1/n_v)$, this sampled vertex is $v$ itself. In that happens, the isolating cut lemma would return the minimal $(s,v)$-mincut, which is exactly $S_v$. This allows us to mark all the $n_v$ vertices that are in $Z$ and appear in $S_v$. So, roughly speaking, we are able to mark at least $n_v$ vertices with probability $1/n_v$ in this case. Of course, we do not know the value of $n_v$, but we try all sampling levels in inverse powers of $2$. We formalize and refine this argument to show that in expectation, we can indeed mark an $\Om(1/\logn)$ fraction of vertices $z\in Z$ with $\mincut(s, z) \le \lambda$.

We now use the \ct algorithm as a ``sieve'' to obtain an \ssc algorithm. We start with $\mincut(s, v)$ for all vertices $v\in V\setminus \{s\}$ tentatively set to the maximum possible edge connectivity (call it $\lambda_{\max}$). Next, we run the \ct algorithm with $\lambda = (1-\e) \lambda_{\max}$. The vertices $v$ that are identified by this algorithm as having $\mincut(s, v) \le \lambda$ drop down to the next level of the hierarchy, while the remaining vertices $v'$ are declared to have $\mincut(s, v') \in ((1-\e)\lambda, \lambda]$. In the next level of the hierarchy, we again invoke the \ct algorithm, but now with $\lambda$ equal to $(1-\e)$ factor of the previous iteration. In this manner, we iteratively continue moving down the hierarchy, cutting the threshold $\lambda$ by a factor of $(1-\e)$ in every step, until the connectivity of all vertices has been determined.

Finally, we come to the problem of obtaining an approximate GH tree. Gomory and Hu's original algorithm uses the following strategy: find an $(s,t)$-mincut for any pair of vertices $s$ and $t$, and recurse on the two sides of the cut in separate subproblems where the other side of the cut is contracted to a single vertex. They used submodularity of cuts to show that contracting one side of an $(s,t)$-mincut does not change the connectivity between vertices on the other side. Moreover, they gave a procedure for combining the two GH trees returned by the recursive calls into a single GH tree at the end of the recursion. Ideally, we would like to use the same algorithm but replace an exact $(s,t)$-mincut with an approximate one. But now, the connectivities in the recursive subproblems are (additively) distorted by the approximation error of the $(s,t)$-mincut. This imposes two additional restrictions. {\bf (a)} First, the values of the $(s,t)$-mincuts identified in the recursive algorithm must now be monotone non-decreasing with depth of the recursion so that the approximation error on a larger $(s,t)$-mincut doesn't get propagated to a smaller $(s',t')$-mincut further down in the recursion. {\bf (b)} Second, the depth of recursion must now be $\polylog(n)$ so that one can control the buildup of approximation error in the recursion by setting the error parameter in a single step to be $\e/\polylog(n)$. Unfortunately, neither of these conditions is met by Gomory and Hu's algorithm. For instance, the recursion depth can be $n-1$ if each $(s,t)$-mincut is a degree cut. The order of $(s,t)$-mincut values in the recursion is also arbitrary and depends on the choice of $s$ and $t$ in each step (which itself is arbitrary).

Let us first consider condition {\bf (a)}. Instead of finding the $(s,t)$-mincut for an arbitrary pair of terminal vertices $s$ and $t$, suppose we found the Steiner mincut on the terminals, i.e., the cut of smallest value that splits the terminals. This would also suffice in terms of the framework since a Steiner mincut is also an $(s,t)$-mincut for some pair $s, t$. But, it brings additional advantages: namely, we get the monotonicity in cut values with recursive depth that we desire. At a high level, this is the idea that we implement: we use the \ct algorithm (with some technical modifications) where we set the threshold $\lambda$ to the value of the Steiner mincut, and identify a partitioning of the terminals where each subset of the partition represents a $(1+\e)$ approximation to the Steiner mincut. 

But, how do we achieve condition {\bf (b)}? Fixing the vertex $s$ in the invocation of the \ssc algorithm, we can identify terminal vertices $v$ that have $\mincut(s, v)\in ((1-\e)\lambda, \lambda]$, where $\lambda$ is the Steiner mincut. But, these approximate Steiner mincuts might be unbalanced in terms of the number of vertices on the two sides of the cut. To understand the problem, suppose there is a single Steiner mincut identified by the \ct algorithm, and this cut is the degree cut of $s$. Then, one subproblem contains all but one vertex in the next round of recursion; consequently, the recursive depth can be high. We overcome this difficulty in two steps. First, we ensure that the only ``large'' subproblem that we recurse on is the one that contains $s$. This can be ensured by sampling $O(\log n)$ different vertices as $s$, which boosts the probability that $s$ is on the larger side of an unbalanced approximate Steiner mincut. This ensures that in the recursion tree, we can only have a large recursive depth along the path containing $s$. Next, we show that even though we are using an approximate method for detemining mincuts, the approximation error only distorts the connectivities in the subproblems not containing $s$. This ensures that the approximation errors can build up only along paths in the recursion tree that have depth $O(\log n)$. Combining these two techniques, we obtain our overall algorithm for an approximate GH tree.

%\section{Preliminaries}
%\input{prelim}

\section{$(1+\e)$-approximate Single Source Min-Cut Algorithm}
\subsection{Preliminaries}
We have already defined the \ssc problem, but for our analysis, we need some more definitions. In particular, we first define a Gomory-Hu Steiner tree and its approximation version.

\BD[Gomory-Hu Steiner tree]
Given a graph $G=(V,E)$ and a set of terminals $U\s V$, the Gomory-Hu Steiner tree is a weighted tree $T$ on the vertices $U$, together with a function $f:V\to U$, such that
 \BI
 \im For all $s,t\in U$, consider the minimum-weight edge $(u,v)$ on the unique $s-t$ path in $T$. Let $U'$ be the vertices of the connected component of $T-(u,v)$ containing $s$.
Then, the set $f\inv(U')\s V$ is an $(s,t)$-mincut, and its value is $w_T(u,v)$.
 \EI
\ED

\BD[Approximate Gomory-Hu Steiner tree]
Given a graph $G=(V,E)$ and a set of terminals $U\s V$, the $(1+\e)$-approximate Gomory-Hu Steiner tree is a weighted tree $T$ on the vertices $U$, together with a function $f:V\to U$, such that
 \BI
 \im For all $s,t\in U$, consider the minimum-weight edge $(u,v)$ on the unique $s-t$ path in $T$. Let $U'$ be the vertices of the connected component of $T-(u,v)$ containing $s$.
Then, the set $f\inv(U')\s V$ is a $(1+\e)$-approximate $(s,t)$-mincut, and its value is $w_T(u,v)$.
 \EI
\ED

\subsection{Algorithms for the \ct and \ssc Problems}

As described earlier, the main tool in our \ssc algorithm is an algorithm for the Cut Threshold (\ct) problem. We first describe a single step of the \ref{thr} algorithm (we call this \ref{step}). 

%\alert{This section starts abruptly. Let us add a text description of what we are going to in this section, and a short outline of the algorithm.} 

\begin{algorithm}
\mylabel{step}{\textsc{CutThresholdStep}}\caption{\ref{step}$(G=(V,E),s,U,W,z)$} 
\begin{algorithmic}[1]
\State Initialize $R^0\gets U$ and $D\gets\emptyset$
\For{$i$ from $0$ to $\lf\lg|U|\rf$}
 \State Compute minimum isolating cuts $\{S^i_v:v\in R^i\}$ on inputs $G$ and $R^i$ \linel{Sv}
 \State Let $D^i$ be the union of $S^i_v\cap U$ over all $v\in R^i\sm \{s\}$ satisfying $w(\pt S^i_v)\le W$ and $|S^i_v\cap U|\le z$\linel{D}
 \State $R^{i+1}\gets$ subsample of $R^i$ where each vertex in $R^i\sm \{s\}$ is sampled independently with probability $1/2$, and $s$ is sampled with probability $1$
\EndFor
%\State $D\gets D^0\cup D^1\cup \cds\cup D^{\lf\lg|U|\rf}$
\State\Return $D^0\cup D^1\cup \cds\cup D^{\lf\lg|U|\rf}$
\end{algorithmic}
\end{algorithm}

We remark that throughout this section, we will always set $z=\infty$, so the constraint $|S^i_v\cap U|\le z$ in \line{D} can be ignored. However, the variable $z$ will play a role in the next section on computing a Gomory-Hu tree.

Let $D=D^0\cup D^1\cup \cds\cup D^{\lf\lg|U|\rf}$ be the union of the sets output by the algorithm. Let $D^*$ be all vertices $v\in U\sm s$ for which there exists an $(s,v)$-mincut of weight at most $W$ whose side containing $v$ has at most $z$ vertices in $U$.  

\BL\leml{step}
$\E[|D\cap D^*|] \ge \Omega(|D^*|/\log|U|)$.
\EL
\BP
%We first prove that $D\s D^*$. Each vertex $u\in D$ belongs to some $S^i_v$ satisfying $w(\pt S^i_v)\le W$ and $|S^i_v\cap U|\le z$. %and $S^i_v\cap U=\{v\}$ for some $v\in U\sm s$. In particular, $\pt S^i_v$ is an $(s,u)$-cut with weight at most $ W$ whose side $S^i_v$ containing $u$ has at most $z$ vertices in $U$, so $u\in D^*$.

\begin{figure}\centering
\includegraphics[scale=1]{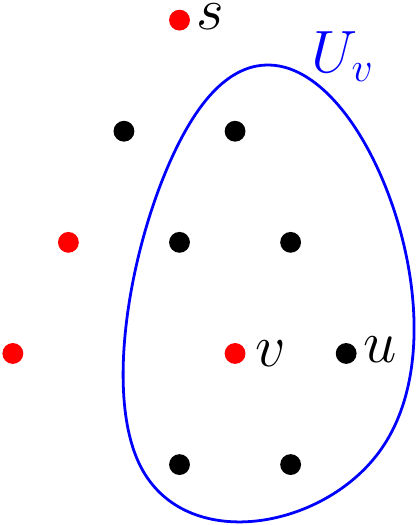}
\caption{Let $i=\lf\lg n_{v}\rf=\lf\lg 7\rf=2$, and let the red vertices be those sampled in $R^2$. Vertex $v$ is active and hits $u$ because $v$ is the only vertex in $U_{v}$ that is red.}\label{fig:hits}
\end{figure}

%It remains to prove that $\E[|D|]\ge\Om(|D^*|/\log|U|)$. 
For each vertex $v\in D^*$, let $S_v$ be the minimal $(v,s)$-mincut, and define $U_v=S_v\cap U$ and $n_v=|U_v|$. We say that a vertex $v\in D^*$ is \emph{active} if $v\in R^i$ for $i=\lf\lg n_{v}\rf$. In addition, if $U_v\cap R^i=\{v\}$, then we say that $v$ \emph{hits} all of the vertices in $U_v$ (including itself); see Figure~\ref{fig:hits}. In particular, in order for $v$ to hit any other vertex, it must be active. For completeness, we say that any vertex in $U\sm D^*$ is not active and does not hit any vertex.

%\alert{The above definitions are fine, but would be easier to read with (a) some intuition behind when you say that a vertex is active and process of hitting other vertices, and (b) a figure illustrating the notation.} 

%\textcolor{blue}{Jason: added a figure.}

To prove that $\E[|D\cap D^*|] \ge \Om(|D^*|/\log|U|)$, we will show that
 \BE
 \im[(a)] each vertex $u$ that is hit is in $D$, 
 \im[(b)] the total number of pairs $(u,v)$ for which $v\in D^*$ hits $u$ is at least $\Om(|D^*|)$ in expectation, and % for some small enough constant $c>0$, and
 \im[(c)] each vertex $u$ is hit by at most $\lf\lg|U|\rf+1$ vertices. %with probability at least $1-\f c{2|U|^2}$ (for the constant $c>0$ in~(b)), each vertex $u$ is hit by at most $O(\log|U|)$ vertices $v\in D^*$.
 \EE

For (a), let $v$ be the vertex that hits $u$, and consider $i=\lfloor\lg n_{v}\rfloor$. We have $U_v\cap R^i=\{v\}$ by assumption, so $\pt S_v$ is a $(v,R^i\sm\{v\})$-cut. On the other hand, we have that $\partial S^i_v$ is a $(v,R^i\setminus\{v\})$-mincut, so in particular, it is a $(v,s)$-cut. It follows that $\partial S_v$ and $\partial S^i_v$ are both $(v,s)$-mincuts and $(v,R^i\setminus v)$-mincuts, and $w(\partial S^i_v)=\mincut(s,v)\le W$. Since $S_v$ is the minimal $(v,s)$-mincut and $S^i_v$ is a $(v,s)$-mincut, we must have $S_v \subseteq S^i_v$. Likewise, since $S_v$ is a $(v,R^i\sm\{v\})$-mincut and $S^i_v$ is the minimal $(v,R^i\setminus\{v\})$-mincut, we also have $S^i_v\subseteq S_v $. It follows that $S_v=S^i_v$. Since $S_v$ is the minimal $(v,s)$-mincut and $v\in D^*$, we must have $|S_v\cap U|\le z$, so in particular, $|S^i_v\cap U|=|S_v\cap U|\le z$. Therefore, the vertex $v$ satisfies all the conditions of line~\ref{line:D}. Moreover, since $u\in U_v\subseteq S_v= S^i_v$, vertex $u$ is added to $D$ in the set $S^i_v\cap U$. 

For (b), for $i=\lf\lg n_{v}\rf$, we have $v\in R^i$ with probability exactly $1/2^i = \Th(1/n_{v})$, and with probability $\Om(1)$, no other vertex in $U_{v}$ joins $R^i$. Therefore, $v$ is active with probability $\Om(1/n_{v})$. Conditioned on $v$ being active, it hits exactly $n_{v}$ many vertices. It follows that $v$ hits $\Om(1)$ vertices in expectation. Summing over all $v\in D^*$ and applying linearity of expectation proves~(b).

For (c), since the isolating cuts $S^i_v$ over $v\in R^i$ are disjoint for each $i$, each vertex is hit at most once on each iteration $i$. Since there are $\lf\lg|U|\rf+1$ many iterations, the property follows. %the number of vertices $v$ that hit vertex $u$ is at most the number of active vertices $v$ for which $v$ is on the path from $u$ to $s$ in $T$. Label these vertices $u=v_1,v_2,\lds,v_\el=s$, ordered by increasing distance from $u$ to $r(v_i)$ in $T$. Each vertex $v_j\in D^*$ is active with probability $\Th(1/n_{r(v_j)})$, which is at most $\Th(1/j)$ since $v_1,\lds,v_j \in U_{r(v_j)}$. Each vertex $v_j\notin D^*$ is never active. Therefore, the expected number of active vertices on the path from $u$ to $s$ is at most $\sum_{j=1}^\el\Th(1/j)=\Th(\ln\el)\le \Th(\ln|U|)$. A standard Chernoff bound shows that with probability at least $1-\f c{2|U|^3}$ for any constant $c>0$, the number of active vertices on the path is indeed $O(\ln|U|)$, where the $O(\cd)$ hides the dependency on $c$. Taking a union bound over all $u\in U$, the probability that this is true for all vertices is at least $1-\f c{2|U|^2}$.

Finally, we show why properties (a) to (c) imply $\E[|D\cap D^*|] \ge \Om(|D^*|/\log|U|)$. By property~(b), the number of times some vertex hits another vertex is $\Om(|D^*|)$ in expectation. Since each vertex is hit at most $O(\log|U|)$ times by property~(c), there are at least $\Om(|D^*|/\log|U|)$ vertices hit in expectation, all of which are included in $D$ by property~(a).
\EP

%The following corollary will be useful in the next section:
%\BC\corl{Dmax}
%Let $D_{\max}$ be the largest set $\bigcup_{v\in R^i} (S^i_v\cap U)$ over all iterations $i$, and let $i_{\max}$ be the corresponding iteration. Then, $\E[|D_{\max}|] \ge \Om(|D^*|/\log^2|U|)$.
%\EC
%\BP
%There are $\lf\lg{U}\rf+1$ iterations in which we add to $D$, so $|D_{\max}|\ge |D|/(\lf\lg|U|\rf+1)$.
%Combining this with $\E[|D|]\ge\Om(|D^*|/\log|U|)$ from \lem{step} proves the claim. 
%\EP

We now use iterate Algorithm~\ref{step} to obtain the \ref{thr} algorithm:

\begin{algorithm}
\mylabel{thr}{\textsc{CutThreshold}}\caption{\ref{thr}$(G=(V,E),s, W)$}
\begin{algorithmic}[1]
\State Initialize $U\gets V$ and $D_{\textup{total}}\gets\emptyset$
\For{$O(\log^2n)$ iterations}
 \State Let $D$ be the union of the sets output by $\ref{step}(G,s,U, W,\infty)$
 \State Update $D_{\textup{total}}\gets D_{\textup{total}}\cup D$ and $U\gets U\sm D$
\EndFor
\State\Return $D_{\textup{total}}$
\end{algorithmic}
\end{algorithm}

\BC\corl{threshold}
W.h.p., the output $D_{\textup{total}}$ of \ref{thr} is exactly all vertices $v\in U\sm \{s\}$ for which the $(s,v)$-mincut has weight at most $ W$. 
\EC
\BP
Let $D^*$ be the target output.
From \line{D} of \ref{step}, it is clear that each vertex $v\in D$ output by the algorithm satisfies $\mincut(s,v)\le W$, so $D_{\textup{total}}\s D^*$. It remains to prove that $D_{\textup{total}}\supseteq D^*$.

By \lem{step}, $|U\cap D^*|$ decreases by $\Om(|D^*|/\log n)$ in expectation. After $O(\log^2n)$ iterations, we have $\E[|U\cap D^*|] \le 1/\poly(n)$, so w.h.p., $U\cap D^*=\emptyset$ and all vertices in $D^*$ are added to $D_{\textup{total}}$.
\EP
In other words, \ref{thr} is an algorithm that fulfills \thm{thr}, restated below.
\Thr*

Finally, we use the \ref{thr} algorithm to design our \ssc algorithm:

\begin{algorithm}
\caption{\textsc{ApproxSSMC}$(G=(V,E),s,\e)$}
\begin{algorithmic}[1]
 \State Initialize bounds: $ w_{\min} \gets$ minimum weight of an edge in $G$, and $ w_{\max}\gets$ maximum weight of an edge %\Comment{Every $(s,v)$-mincut has weight in $[ w_{\min}, w_{\max}]$}
 \For{all integers $j\ge0$ s.t.\ $(1+\e)^j w_{\min} \in [ w_{\min},(1+\e) n w_{\max}]$}
 \State $ W_j\gets (1+\e)^j w_{\min}$
 \State $D_j\gets \ref{thr}(G,s, W)$
\EndFor
\State For each vertex $v\in V$, take the largest $D_i$ containing $v$, and set $\tilde\la(v)\gets  W_i$
\State\Return $\tilde\la:V\to \R$
\end{algorithmic}
\end{algorithm}

\BL
W.h.p., the output $\tilde\la$ of \textsc{ApproxSSMC} satisfies $\mincut(s,v) \le \tilde\la(v) \le (1+\e)\mincut(s,v)$.
\EL
\BP
For all $v\in V\sm\{s\}$, we have $w_{\min}\le\mincut(s,v)\le w(\pt(\{s\}))\le nw_{\max}$, so there is an integer $j$ with $ W_j\in[\mincut(s,v),(1+\e)\mincut(s,v))$. The lemma follows from \cor{threshold} applied to this $j$.
%Follows from \cor{threshold} and the fact that for all $v\in V$, there is an integer $i$ with $ W_i\in[\mincut(s,v),(1+\e)\mincut(s,v))$. 
\EP

We have therefore proved \thm{ssmc}, restated below.
\SSMC*

\section{Approximate Gomory-Hu Steiner Tree}
\subsection{Unweighted Graphs}

Let $\e>0$ be a fixed parameter throughout the recursive algorithm. We present our approximate Gomory-Hu Steiner tree algorithm in \ref{approxGH} below.
See Figure~\ref{fig:recursion} for a visual guide to the algorithm.

At a high level, the algorithm applies divide-and-conquer by cutting the graph along sets $S^i_v$ computed by \ref{step}, applying recursion to each piece, and stitching the recursive Gomory-Hu trees together in the same way as the standard recursive Gomory-Hu tree construction. To avoid complications, we only select sets $S^i_v$ from a \emph{single} level $i\in\{0,1,2\lds,\lf\lg|U|\rf\}$, which are guaranteed to be vertex-disjoint. Furthermore, instead of selecting all sets $\{S^i_v:v\in R^i\}$, we only select those for which $|S^i_v\cap U|\le|U|/2$; this allows us to bound the recursion depth. By choosing the source $s\in U$ at \emph{random}, we guarantee that in expectation, we do not exclude too many sets $S^i_v$. The chosen sets partition the graph into disjoint sets of vertices (including the set of vertices outside of any chosen set $S^i_v$). We split the graph along this partition a similar way to the standard Gomory-Hu tree construction: for each set in the partition, contract all other vertices into a single vertex and recursively compute the Gomory-Hu Steiner tree of the contracted graph. This gives us a collection of Gomory-Hu Steiner trees, which we then stitch together into a single Gomory-Hu Steiner tree in the standard way.

%\alert{DP: Should we add a description of the algorithm in text?} \textcolor{blue}{JL: wrote a description.}

%\alert{DP: To simplify notation, how about limiting the CutThresholdStep subroutine to a single invocation of the isolating cut lemma, i.e., a single sampling rate? The loop that runs over different sampling rates can be brought outside in CutThreshold. Here, we can again repeat this loop, thereby giving a clearer description of the notation indexed by $j$.}\textcolor{blue}{JL: I decided against it, because \ref{approxGH} actually needs the sets $S^i_v$ in order to filter out the sets with $|S^i_v\cap U|>|U|/2$. So it might be better to just white-box call \ref{step}, extracting out the intermediate variables $S^i_v$, instead of making \ref{step} return them as well.}

\begin{figure}\centering
\includegraphics[scale=.5]{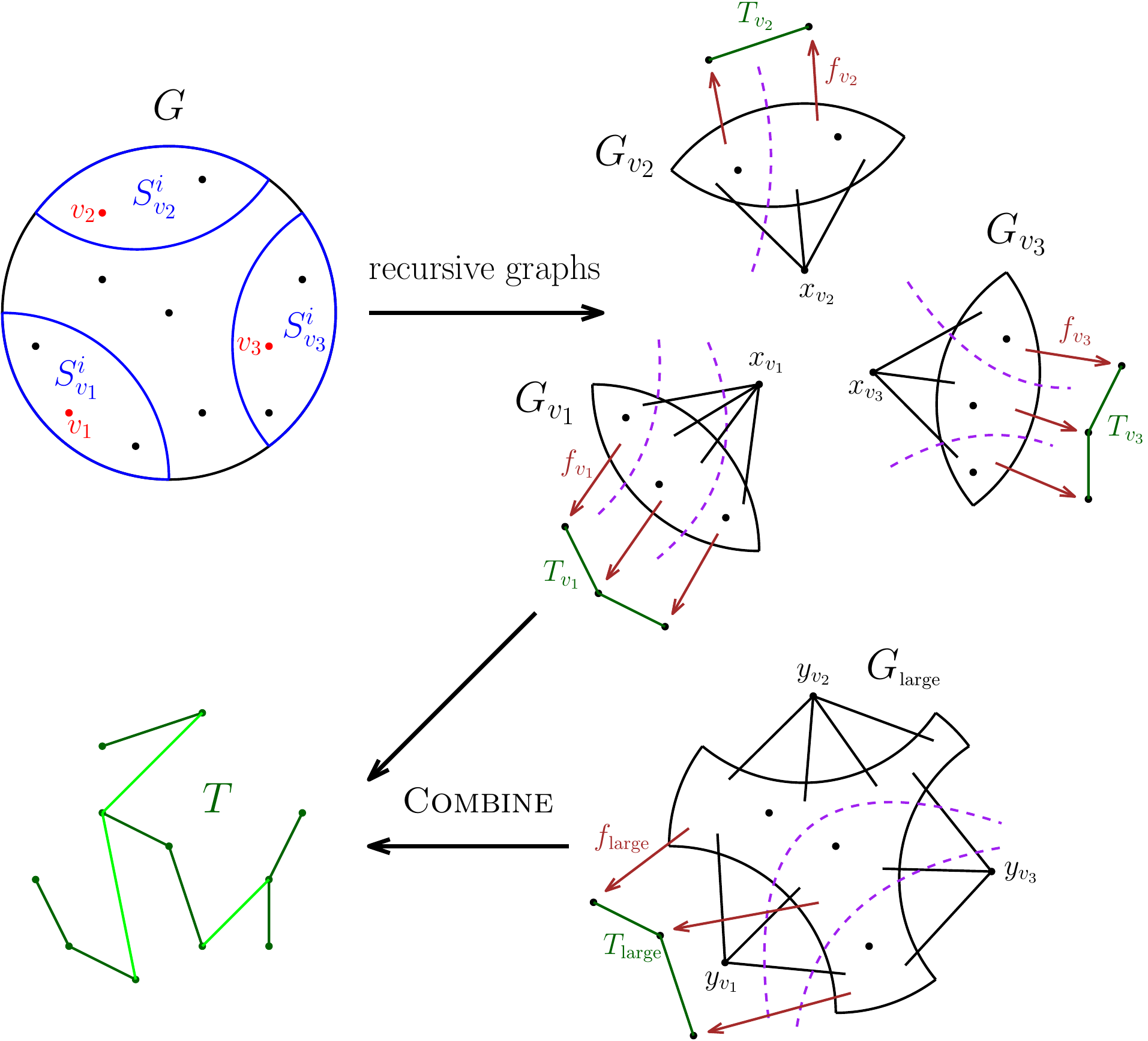}
\caption{Recursive construction of $G_\lar$ and $G_v$ for $v\in R^i_\sma$. Here, $R^i_\sma=\{v_1,v_2,v_3\}$, denoted by red vertices on the top left. The dotted blue curves on the right mark the boundaries of the regions $f_{v_i}\inv(u):u\in U_{v_i}$ and $f_{v_\lar}\inv(u):u\in U_\lar$. The light green edges on the bottom left are the edges $(f_{v_i}(x_{v_i}),f_\lar(y_{v_i}))$ added on \line{combine-T} of \ref{combine}.}\label{fig:recursion}
\end{figure}

\begin{algorithm}
\mylabel{approxGH}{\textsc{ApproxSteinerGHTree}}\caption{\ref{approxGH}$(G=(V,E),U)$} 
\begin{algorithmic}[1]
\State $\la\gets $ global Steiner mincut of $G$ with terminals $U$ %\Comment{Sparsify if necessary}
%\If{}
% \State $G'\gets\textsc{Sparsify}(G,)$
% \State\Return \ref{approxGH}$(G=(V,E),U,\e)$
%\EndIf

\State $s\gets$ uniformly random vertex in $U$
\State Call $\ref{step}(G,s,U,(1+\e)\la,|U|/2)$, and let $R^j$ and $S^j_v:v\in R^j$ ($0\le j\le\lg|U|$) be the intermediate variables in the algorithm\linel{thr}
\State Let $i\in\{0,1,\lds,\lf\lg|U|\rf\}$ be the iteration maximizing $\big|\bigcup_{v\in R^i} (S^i_v\cap U)\big|$ \linel{i}

\ \linel{max}
\For{each $v\in  R^i$} \Comment{Construct recursive graphs and apply recursion}
 \State Let $G_v$ be the graph $G$ with vertices $V\sm S^i_v$ contracted to a single vertex $x_v$ \Comment{$S^i_v$ are disjoint}
 \State Let $U_v\gets S^i_v\cap U$
 \State $(T_v,f_v)\gets\ref{approxGH}(G_v,U_v)$
\EndFor
\State Let $G_\lar$ be the graph $G$ with (disjoint) vertex sets $S^i_v$ contracted to single vertices $y_v$ for all $v\in R^i$
\State Let $U_\lar\gets U\sm\bigcup_{v\in R^i}(S^i_v\cap U)$
\State $(T_\lar,f_\lar)\gets\ref{approxGH}(G_\lar,U_\lar)$

\
\State Combine $(T_\lar,f_\lar)$ and $\{(T_v,f_v):v\in R^i\}$ into $(T,f)$ according to \ref{combine}%$((T_\lar,f_\lar),\{(T_v,f_v):v\in R^i)\}$
%\State Construct $T$ by starting with the disjoint union $T_\lar\cup\bigcup_{v\in R^i}T_v$ and, for each $v\in R^i$, adding an edge between $f_v(x_v)\in U_v$ and $f_\lar(y_v)\in U_\lar$ of weight $w(\pt_GS^i_v)$ 

%\Comment{Combine Gomory-Hu Steiner trees together} \linel{combine-T}
%\State Construct $f:V\to U$ by $f(v')=f_\lar(v')$ if $v'\in U_\lar$ and $f(v')=f_v(v')$ if $v'\in U_v$ for some $v\in R^i$\linel{combine-f}
\State\Return $(T,f)$

\end{algorithmic}
\end{algorithm}

\begin{algorithm}
\mylabel{combine}{\textsc{Combine}}\caption{\ref{combine}$((T_\lar,f_\lar),\{(T_v,f_v): v\in R^i\} )$} 
\begin{algorithmic}[1]
\State Construct $T$ by starting with the disjoint union $T_\lar\cup\bigcup_{v\in R^i}T_v$ and, for each $v\in R^i$, adding an edge between $f_v(x_v)\in U_v$ and $f_\lar(y_v)\in U_\lar$ of weight $w(\pt_GS^i_v)$\linel{combine-T}
\State Construct $f:V\to U$ by $f(v')=f_\lar(v')$ if $v'\in U_\lar$ and $f(v')=f_v(v')$ if $v'\in U_v$ for some $v\in R^i$\linel{combine-f}
\State\Return $(T,f)$
\end{algorithmic}
\end{algorithm}

%\alert{DP: I think a figure will help in explaining how the recursion proceeds, both in terms of what subproblems are generated and how the reconstruction happens.}

\subsection{Approximation}

Since the approximation factors can potentially add up down the recursion tree, we need to bound the depth of the recursive algorithm. Here, there are two types of recursion: the recursive calls $(G_v,U_v)$, and the single call $(G_\lar,U_\lar)$.  Taking a branch down $(G_v,U_v)$ is easy: since $|U_v|\le|U|/2$, the algorithm can travel down such a branch at most $\lg|U|$ times. The difficult part is in bounding the number of branches down $(G_\lar,U_\lar)$. It turns out that after $\pl(n)$ consecutive branches down $(G_\lar,U_\lar)$, the Steiner mincut increases by factor $(1+\e)$, w.h.p.; we elaborate on this insight in \sec{running}, which concerns the running time. Since the Steiner mincut can never decrease down any recursive branch, it can increase by factor $(1+\e)$ at most $\e\inv\pl(n)\log\De$ times. Thus, we have a bound of $\e\inv\pl(n)\log\De$ on the recursion depth, w.h.p.

This depth bound alone is not enough for the following reason: if the approximation factor increase by $(1+\e)$ along each recursive branch, then the total approximation becomes $(1+\e)^{\e\inv\pl(n)\log\De}$, which is no good because the $(1+\e)$ and $\e\inv$ cancel each other. Here, our key insight is that actually, the approximation factor does not distort \emph{at all} down $(G_\lar,U_\lar)$. It may increase by factor $(1+\e)$ down any $(G_v,U_v)$, but this can only happen $\lg|U|$ times, giving us an approximation factor of $(1+\e)^{\lg|U|}$, which is fine because we can always retroactively replace $\e$ with $\Th(\e/\lg|U|)$ to obtain the desired $(1+\e)$.

The lemma below formalizes our insight that approximation factors are preserved down the branch $(G_\lar,U_\lar)$.

\BL\leml{large}
For any distinct vertices $p,q\in U_\lar$, we have $\mincut_{G_\lar}(p,q) = \mincut_G(p,q)$.
\EL
\BP
Since $G_\lar$ is a contraction of $G$, we have $\mincut_{G_\lar}(p,q) \ge \mincut_G(p,q)$. To show the reverse inequality, fix any $(p,q)$-mincut in $G$, and let $S$ be one side of the mincut. We show that for each $v\in  R^i$, either $S^i_v\s S$ or $S^i_v\s V\sm S$. Assuming this, the cut $\pt S$ stays intact when the sets $S^i_v$ are contracted to form $G_\lar$, so $\mincut_{G_\lar}(p,q) \le w(\pt S) = \mincut_G(p,q)$.

Consider any $v\in R^i$, and suppose first that $v\in S$. Then, $S^i_v\cap S$ is still a $(v,R^i\sm v)$-cut, and $S^i_v\cup S$ is still a $(p,q)$-cut. By the submodularity of cuts,
\[ w(\pt_GS^i_v) + w(\pt_GS) \ge w(\pt_G(S^i_v\cup S)) + w(\pt_G(S^i_v\cap S)). \]
In particular, $S^i_v\cap S$ must be a minimum $(v,R^i\sm v)$-cut. Since $S^i_v$ is the minimal $(v,R^i\sm v)$-mincut, it follows that $S^i_v\cap S = S^i_v$, or equivalently, $S^i_v\s S$.

Suppose now that $v\notin S$. In this case, we can swap $p$ and $q$, and swap $S$ and $V\sm S$, and repeat the above argument to get $S^i_v\s V\sm S$.
\EP

Similarly, the lemma below says that approximation factors distort by at most $(1+\e)$ down a $(G_v,U_v)$ branch.

\BL\leml{small}
For any $v\in  R^i$ and any distinct vertices $p,q\in U_v$, we have $\mincut_G(p,q)\le\mincut_{G_v}(p,q)\le(1+\e)\mincut_G(p,q)$.
\EL
\BP
The lower bound $\mincut_G(p,q)\le\mincut_{G_v}(p,q)$ holds because $G_v$ is a contraction of $G$, so we focus on the upper bound.
Fix any $(p,q)$-mincut in $G$, and let $S$ be the side of the mincut not containing $s$ (recall that $s\in U$ and $s\notin S^i_v$). Since $S^i_v\cup S$ is a $(p, s)$-cut (it is also a $(q, s)$-cut), it is in particular a Steiner cut for terminals $U$, so $w(S^i_v\cup S)\ge\la$. Also, $w(S^i_v)\le(1+\e)\la$ by the choice of the threshold $(1+\e)\la$ (\line{thr}). Together with the submodularity of cuts, we obtain
\BAL (1+\e)\la+w(\pt_GS) &\ge w(\pt_GS^i_v) + w(\pt_GS) \\&\ge w(\pt_G(S^i_v\cup S)) + w(\pt_G(S^i_v\cap S)) \\&\ge \la + w(\pt_G(S^i_v\cap S)) .\EAL
The set $S^i_v\cap S$ stays intact under the contraction from $G$ to $G_v$, so $w(\pt_{G_v}(S^i_v\cap S))=w(\pt_G(S^i_v\cap S))$. Therefore,
\BAL \mincut_{G_v}(p,q)&\le w(\pt_{G_v}(S^i_v\cap S))\\&=w(\pt_G(S^i_v\cap S)) \\&\le w(\pt_GS)+\e\la \\&\le \mincut_G(p,q) + \e\,\mincut_G(p,q) ,\EAL
as promised.
\EP

Finally, the lemma below determines our final approximation factor.

\BL\leml{approx}
\ref{approxGH}$(G=(V,E),U)$ outputs a $(1+\e)^{\lg|U|}$-approximate Gomory-Hu Steiner tree.
\EL
\BP
We apply induction on $|U|$. 
Since $|U_v|\le|U|/2$ for all $v\in R^i$, by induction, the recursive outputs $(T_v,f_v)$ are Gomory-Hu Steiner trees with approximation $(1+\e)^{\lg|U_v|}\le(1+\e)^{\lg|U|-1}$.  By definition, this means that for all $s,t\in U_v$ and the minimum-weight edge $(u,u')$ on the $s$--$t$ path in $T_v$, letting $U'_v\s U_v$ be the vertices of the connected component of $T_v-(u,u')$ containing $s$, we have that $f\inv_v(U'_v)$ is a $(1+\e)^{\lg|U|-1}$-approximate $(s,t)$-mincut in $G_v$ with value $w_T(u,u')$. Define $U'\s U$ as the vertices of the connected component of $T-(u,u')$ containing $s$. By construction of $(T,f)$ (lines~\ref{line:combine-T}~and~\ref{line:combine-f}), the set $f\inv(U')$ is simply $f\inv_v(U'_v)$ with the vertex $x_v$ replaced by $V\sm S^i_v$ in the case that $x_v\in f\inv(U')$. Since $G_v$ is simply $G$ with all vertices $V\sm S^i_v$ contracted to $x_v$, we conclude that $w_{G_v}(\pt f\inv_v(U'_v)) = w_G(\pt f\inv(U'))$. By \lem{small}, the values $\mincut_G(s,t)$ and $\mincut_{G_v}(s,t)$ are within factor $(1+\e)$ of each other, so $w_G(\pt f\inv(U'))$ approximates the $(s,t)$-mincut in $G$ to a factor $(1+\e)\cd(1+\e)^{\lg|U|-1} = (1+\e)^{\lg|U|}$. In other words, the Gomory-Hu Steiner tree condition for $(T,f)$ is satisfied for all $s,t\in U_v$ for some $v\in R^i$.

By induction, the recursive output $(T_\lar,f_\lar)$ is a Gomory-Hu Steiner tree with approximation $(1+\e)^{\lg|U_\lar|}\le(1+\e)^{\lg|U|}$. Again, consider $s,t\in U_\lar$ and the minimum-weight edge $(u,u')$ on the $s$--$t$ path in $T_\lar$, and let $U'_\lar\s U_\lar$ be the vertices of the connected component of $T_\lar-(u,u')$ containing $s$. Define $U'\s U$ as the vertices of the connected component of $T-(u,u')$ containing $s$. By a similar argument, we have $w_{G_\lar}(\pt f\inv_\lar(U'_\lar)) = w_G(\pt f\inv(U'))$. By \lem{large}, we also have $\mincut_G(s,t)=\mincut_{G_\lar}(s,t)$, so $w_G(\pt f\inv(U'))$ is a $(1+\e)^{\lg|U|}$-approximate $(s,t)$-mincut in $G$, fulfilling the Gomory-Hu Steiner tree condition for $(T,f)$ in the case $s,t\in U_\lar$.

There are two remaining cases: $s\in U_v$ and $t\in U_{v'}$ for distinct $v,v'\in R^i$, and $s\in U_v$ and $t\in U_\lar$; we treat both cases simultaneously. Since $G$ has Steiner mincut $\la$, each of the contracted graphs $G_\lar$ and $G_v$ has Steiner mincut at least $\la$. Since all edges on the approximate Gomory-Hu Steiner tree correspond to actual cuts in the graph, every edge in $T_v$ and $T_\lar$ has weight at least $\la$. By construction, the $s$--$t$ path in $T$ has at least one edge of the form $(f_v(x_v),f_\lar(y_v))$, added on \line{combine-T}; this edge has weight $w(\pt_GS^i_v)\le(1+\e)\la$. Therefore, the minimum-weight edge on the $s$--$t$ path in $T$ has weight at least $\la$ and at most $(1+\e)\la$; in particular, it is a $(1+\e)$-approximation of $\mincut_G(s,t)$. If the edge is of the form $(f_v(x_v),f_\lar(y_v))$, then by construction, the relevant set $f\inv(U')$ is exactly $S^i_v$, which is a $(1+\e)$-approximate $(s,t)$-mincut in $G$. If the edge is in $T_\lar$ or $T_v$ or $T_{v'}$, then we can apply the same arguments used previously. %\alert{(sketch) Can only go to $G_v$ at most $\lg|U|$ times, picks up $(1+\e)$ factor each time}
\EP

\subsection{Running Time Bound}\secl{running}

In order for a recursive algorithm to be efficient, it must make substantial progress on each of its recursive calls, which can then be used to bound its depth. For each recursive call $(G_v,U_v,\e)$, we have $|U_v|\le|U|/2$ by construction, so we can set our measure of progress to be $|U|$, the number of terminals, which halves upon each recursive call.
However, progress on $(G_\lar,U_\lar,\e)$ is unclear; in particular, it is possible for $|U_\lar|$ to be very close to $|U|$ with probability $1$. For $G_\lar$, we define the following alternative measure of progress. Let $P(G,U,W)$ be the set of unordered pairs of distinct vertices whose mincut is at most $W$:
\[ P(G,U,W) = \bigg\{ \{u,v\}\in\bn U2:\mincut_G(u,v)\le W \bigg\} .\]
In particular, we will consider its size $|P(G,U,W)|$, and show the following expected reduction:

\begin{restatable}{lemma}{LemD}\leml{P}
For any $W\le(1+\e)\la$, over the random selection of $s$ and the randomness in \ref{step}, we have
\[ \E[|P(G_\lar,U_\lar,W)|] \le \lp1-\Om\lp\f1{\log^2n}\rp\rp|P(G,U,W)| .\]
\end{restatable}

Before we prove \lem{P}, we show how it implies progress on the recursive call for $G_\lar$.
\BC\corl{mincut-increase}
Let $\la_0$ be the global Steiner mincut of $G$.
W.h.p., after $\Om(\log^3n)$ recursive calls along $G_\lar$ (replacing $G\gets G_\lar$ each time), the global Steiner mincut of $G$ is at least $(1+\e)\la_0$ (where $\la_0$ is still the global Steiner mincut of the initial graph).
\EC
\BP
Let $W=(1+\e)\la_0$.
Initially, we trivially have $|P(G,U,W)|\le\bn{|U|}2$. The global Steiner mincut can only increase in the recursive calls, since $G_\lar$ is always a contraction of $G$, so we always have $W\le(1+\e)\la$ for the current global Steiner mincut $\la$. By \lem{P}, the value $|P(G,U,W)|$ drops by factor $1-\Om(\f1{\log^2n})$ in expectation on each recursive call, so after $\Om(\log^3n)$ calls, we have
\[ \E[|P(G,U,W)|]\le\bn{|U|}2\cd\lp1-\Om\lp\f1{\log^2n}\rp\rp^{\Om(\log^3n)}\le\f1{\poly(n)} .\]
In other words, w.h.p., we have $|P(G,U,W)|=0$ at the end, or equivalently, the Steiner mincut of $G$ is at least $(1+\e)\la$.
\EP

Combining both recursive measures of progress together, we obtain the following bound on the recursion depth:
\BL\leml{depth}
%Let $w_{\min}=\min_{s,t\in U}\mincut(s,t)$ and $w_{\max}=\max_{s,t\in U}\mincut(s,t)$ be the minimum and maximum mincuts between two terminals.
Let $w_{\min}$ and $w_{\max}$ be the minimum weight and maximum weight of any edge in $G$.
W.h.p., the depth of the recursion tree of \ref{approxGH} is $O(\e\inv\log^3n\log(n\De))$.
\EL
\BP
For any $\Th(\log^3n)$ successive recursive calls down the recursion tree, either one call was on a graph $G_v$, or $\Th(\log^3n)$ of them were on the graph $G_\lar$. In the former case, $|U|$ drops by half, so it can happen $O(\logn)$ times total. In the latter case, by \cor{mincut-increase}, the global Steiner mincut increases by factor $(1+\e)$. Let $w_{\min}$ and $w_{\max}$ be the minimum and maximum weights in $G$, so that $\De=w_{\max}/w_{\min}$. Note that for any recursive instance $(G',U')$ and any $s,t\in U'$, we have $w_{\min}\le\mincut_{G'}(s,t)\le w(\pt(\{s\}))\le nw_{\max}$, so the global Steiner mincut of $(G',U')$ is always in the range $[w_{\min},nw_{\max}]$. It follows that calling $G_\lar$ can happen $O(\e\inv\log(nw_{\max}/w_{\min}))$ times, hence the bound.
\EP

We state the next theorem for \emph{unweighted} graphs only. For weighted graphs, there is no nice bound on the number of new edges created throughout the algorithm, and therefore no easy bound on the overall running time. In the next section, we introduce a graph sparsification step to handle this issue.

\BL\leml{runtime}
For an \emph{unweighted} graph $G=(V,E)$, and terminals $U\s V$,\linebreak $\ref{approxGH}(G,V,\e)$ takes time $\tO(m\e\inv)$ plus calls to max-flow on instances with a total of $\tO(n\e\inv)$ vertices and $\tO(m\e\inv)$ edges.% many calls to max-flow on $O(n)$-vertex, $O(m)$-edge graphs.
\EL
\BP
For a given recursion level, consider the instances $\{ (G_i,U_i,W_i)\}$ across that level. By construction, the terminals $U_i$ partition $U$. Moreover, the total number of vertices over all $G_i$ is at most $n+2(|U|-1)=O(n)$ since each branch creates $2$ new vertices and there are at most $|U|-1$ branches. The total number of new edges created is at most the sum of weights of the edges in the final $(1+\e)$-approximate Gomory-Hu Steiner tree. For an unweighted graph, this is $O(m)$ by the following well-known argument. Root the Gomory-Hu Steiner tree $T$ at any vertex $r\in U$; for any $v\in U\sm r$ with parent $u$, the cut $\pt\{v\}$ in $G$ is a $(u,v)$-cut of value $\deg(v)$, so $w_T(u,v)\le\deg(v)$. Overall, the sum of the edge weights in $T$ is at most $\sum_{v\in U}\deg(v)\le2m$.

Therefore, there are $O(n)$ vertices and $O(m)$ edges in each recursion level. By \lem{depth}, there are $O(\e\inv\log^4n)$ levels (since $\De=1$ for an unweighted graph), for a total of $\tO(n\e\inv)$ vertices and $\tO(m\e\inv)$ edges. In particular, the instances to the max-flow calls have $\tO(n\e\inv)$ vertices and $\tO(m\e\inv)$ edges in total.
\EP

Combining \Cref{lem:approx,lem:runtime} and resetting $\e\gets\Th(\e/\logn)$, we obtain \thm{approx-u}, restated below.
\ApproxU*

Finally, we prove \lem{P}, restated below.
\LemD*
\BP
Let $D^*$ be all vertices $v\in U\sm s$ for which there exists an $(s,v)$-cut of weight at most $W$ whose side containing $v$ has at most $|U|/2$ vertices in $U$. Define $D = \bigcup_{j=0}^{\lf\lg|U|\rf}\bigcup_{v\in R^i} (S^i_v\cap U)$.
Let $P_{\text{ordered}}(G,U,W)$ be the set of ordered pairs $(u,v):u,v\in V$ for which there exists an $(u,v)$-mincut of weight at most $W$ with at most $|U|/2$ vertices in $U$ on the side $S(u,v)\s V$ containing $u$. %Observe that $|P_{\text{ordered}}(G,U,W)| \ge |P(G,U,W)|$, since for each pair $(u,v)$.
%For each ordered pair $(u,v)$ for which $\{u,v\}\in P(G,U,W)$, For each unordered pair $\{u,v\}\in P(G,U,W)$, consider the $(u,v)$-mincut of weight at most $W$. Let $S(\{u,v\})\s V$ be the side of the mincut with the smaller number of vertices in $U$, with a tie broken arbitrarily, and let $f(\{u,v\})\in\{u,v\}$ be whichever vertex ($u$ or $v$) is in $S(\{u,v\})$.
We now state and prove the following four properties:

\BE
\im[(a)] For all $u,v\in U$, $\{u,v\}\in P(G,U,W)$ if and only if either $(u,v)\in P_{\text{ordered}}(G,U,W)$ or $(v,u)\in P_{\text{ordered}}(G,U,W)$ (or both).
\im[(b)] For each pair $(u,v)\in P_{\text{ordered}}(G,U,W)$, we have $u\in D^*$ with probability at least $1/2$,
\im[(c)] For each $u\in D^*$, there are at least $|U|/2$ vertices $v\in U$ for which $(u,v)\in P_{\text{ordered}}(G,U,W)$.
\im[(d)] Over the randomness in \ref{step}$(G,U,(1+\e)\la)$, $\E[|D|]\ge\Om(|D^*|/\log|U|)$.
\EE

Property (a) follows by definition.
Property~(b) follows from the fact that $u\in D^*$ whenever $s\notin S(u,v)$, which happens with probability at least $1/2$. 
Property~(c) follows because any vertex $v\in U\sm S(u,v)$ satisfies $(u,v)\in P_{\text{ordered}}(G,U,W)$, of which there are at least $|U|/2$. Property~(d) follows from \lem{step} applied on  $\ref{step}(G,U,W,|U|/2)$, and then observing that even though we actually call $\ref{step}(G,U,(1+\e)\la,|U|/2)$, the set $D$ can only get larger if the weight parameter is increased from $W$ to $(1+\e)\la$.

With properties (a) to (d) in hand, we now finish the proof of \lem{P}. Consider the iteration $i$ maximizing the size of $D^i := \bigcup_{v\in R^i} (S^i_v\cap U)$ (\line{max}), so that $|D^i|\ge|D|/(\lf\lg|U|\rf+1)$. For any vertex $u\in D^i$, all pairs $(u,v)\in P_{\text{ordered}}(G,U,W)$ (over all $v\in U$) disappear from $P_{\text{ordered}}(G,U,W)$, which is at least $|U|/2$ many by (c). In other words, 
\BAL & |P_{\text{ordered}}(G,U,W)\sm P_{\text{ordered}}(G_\lar,U_\lar,W)| \ge \f{|U|}2|D^i| \ge\Om\lp\f{|U|\cd |D|}{\log|U|}\rp   .\EAL
Taking expectations and applying (d), 
\BAL & \E[|P_{\text{ordered}}(G,U,W)\sm P_{\text{ordered}}(G_\lar,U_\lar,W)|] \ge\Om\lp\f{|U|\cd\E[|D|]}{\log|U|}\rp    \ge \Om\lp\f{|U|\cd|D^*|}{\log^2|U|}\rp  .\EAL
Moreover,
\[ |U|\cd|D^*| \ge \E\big[\big| \{(u,v): u\in D^*\} \big|\big] \ge \f12|P_{\text{ordered}}(G,U,W)|,  \]
where the second inequality follows by (b). Putting everything together, we obtain
\BAL &\E[|P_{\text{ordered}}(G,U,W)\sm P_{\text{ordered}}(G_\lar,U_\lar,W)|] \ge \Om\lp\f{|P_{\text{ordered}}(G,U,W)|}{\log^2|U|}\rp   .\EAL
Finally, applying (a) gives
\BG \E[|P(G,U,W) \sm P(G_\lar,U_\lar,W)|] \ge \Om\lp\f{|P(G,U,W)|}{\log^2|U|}\rp .\nonumber%\eqnl{P}
\EG
Finally, we have $P(G_\lar,U_\lar,W) \s P(G,U,W)$ since the $(u,v)$-mincut for $u,v\in U_\lar$ can only increase in $G_\lar$ due to $G_\lar$ being a contraction of $G$ (in fact it says the same by \lem{large}). Therefore,
\BAL &|P(G,U,W)| - |P(G_\lar,U_\lar,W)| = |P(G,U,W) \sm P(G_\lar,U_\lar,W)| ,\EAL
and combining with the bound on  $\E[|P(G,U,W) \sm P(G_\lar,U_\lar,W)|]$ concludes the proof.
\EP

\subsection{Weighted Graphs}

%Include a sparsification step to force linear number of edges on each recursive instance

For weighted graphs, we cannot easily bound the total size of the recursive instances. Instead, to keep the sizes of the instances small, we sparsify the recursive instances to have roughly the same number of edges and vertices. By the proof of \lem{runtime}, the total number of vertices over all instances in a given recursion level is at most $n+2(|U|-1)=O(n)$. Therefore, if each such instance is sparsified, the total number of edges becomes $\tO(n)$, and the algorithm is efficient.

It turns out we only need to re-sparsify the graph in two cases: when we branch down to a graph $G_v$ (and not $G_\lar$), and when the mincut $\la$ increases by a constant factor, say $2$. The former can happen at most $O(\logn)$ times down any recursion branch, since $|U|$ decreases by a factor $2$ each time, and the latter occurs $O(\log(n\De))$ times down any branch. Each time, we sparsify up to factor $1+\Th(\e/\log(n\De))$, so that the total error along any branch is $1+\Th(\e)$. 

We now formalize our arguments. We begin with the specification routine due to Benczur and Karger~\cite{BenczurKarger1996}.

%\BD[Strength]
%Given a graph $G=(V,E)$, the \emph{strength} of an edge $e\in E$ is the maximum global mincut of any vertex-induced subgraph of $G$ that contains $e$.
%\ED
%\BT\thml{str}
%Given a weighted, undirected graph $G=(V,E)$, there exists a deterministic algorithm in $\tO(m)$ time that computes values $k'_e:e\in E$ such that
% \BE
% \im For each edge $e\in E$, the strength of $e$ is at least $k'_e$, and 
% \im $\sum_{e\in E}k'_e = O(n)$.
% \EE
%\ET
\BT\thml{sparsify}
Given a weighted, undirected graph $G$, and parameters $\e,\de>0$, there is a randomized algorithm that with probability at least $1-\de$ outputs a $(1+\e)$-approximate sparsifier of $G$ with $O(n\e^{-2}\log(n/\de))$ edges. %\alert{DP: You omitted the dependence on $\e$ in the size. Should we keep it?}
%and let $k'_e:e\in E$ be values satisfying the two conditions in \thm{str}. Let $H$ be a weighted, undirected graph obtained by independently sampling each edge $e\in E$ with probability $p_e=\Th(\f{\log n}{\e^2k'_e})$ and (if sampled) weighting the sampled edge by $w_e/p_e$. Then w.h.p., $H$ is a $(1+\e)$-approximate sparsifier of $G$ with $O(n\logn)$ edges.
\ET

We now derive approximation and running time bounds.
\BT
Suppose that the recursive algorithm \ref{approxGH} sparsifies the input in the following three cases, using \thm{sparsify} with the same parameter $\e$ and the parameter $\de=1/\poly(n)$:
 \BE
 \im The instance was the original input, or
 \im The instance was obtained from calling $(G_v,U_v)$, or
 \im The instance was obtained from calling $(G_\lar,U_\lar)$, and the Steiner mincut increased by a factor of at least $2$ since the last sparsification.
 \EE
Then w.h.p., the algorithm outputs a $(1+\e)^{O(\log(n\De))}$-approximate Gomory-Hu Steiner tree and takes $\tO(m)$ time plus calls to maxflow on instances with a total of $\tO(n\e\inv\log\De)$ vertices and $\tO(n\e\inv\log\De)$ edges.
\ET
\BP
We first argue about the approximation factor. Along any branch of the recursion tree, there is at most one sparsification step of type~(1), at most $O(\logn)$ sparsification steps of type~(2), and at most $O(\log(n\De))$ sparsification steps of type~(3). Each sparsification distorts the pairwise mincuts by a $(1+\e)$ factor, so the total distortion is $(1+\e)^{O(\log(n\De))}$.

Next, we consider the running time. The recursion tree can be broken into chains of recursive $G_\lar$ calls, so that each chain begins with either the original instance or some intermediate $G_v$ call, which is sparsified by either~(1)~or~(2). Fix a chain, and let $n'$ be the number of vertices at the start of the chain, so that the number of edges is $O(n'\log n)$. Within each chain, the number of vertices can only decrease down the chain. After each sparsification, many sparsifications of type~(2), and between two consecutive sparsifications, the number of edges can only decrease down the chain since the graph can only contract. It follows that each instance in the chain has at most $n'$ vertices and $O(n'\e^{-2}\logn)$ edges. %\alert{DP: You are omitting $\e$ here in the size. Should we keep it?} 
By \lem{depth}, each chain has length $O(\e\inv\log^3n\log(n\De))$, so the total number of vertices and edges in the chain is $\tO(n'\e^{-3}\log\De)$. Imagine charging these vertices and edges to the $n'$ vertices at the root of the chain.
In other words, to bound the total number of edges in the recursion tree, it suffices to bound the total number of vertices in the original instance and in intermediate $G_v$ calls. 

In the recursion tree, there are $n$ original vertices and at most $2(|U|-1)$ new vertices, since each branch creates $2$ new vertices and there are at most $|U|-1$ branches. Each vertex joins $O(\logn)$ many $G_v$ calls, since every time a vertex joins one, the number of terminals drops by half; note that a vertex is never duplicated in the recursion tree. It follows that there are $O(n\logn)$ many vertices in intermediate $G_v$ calls, along with the $n$ vertices in the original instance. Hence, from our charging scheme, we conclude that there are a total of $\tO(n\e^{-3}\log\De)$ vertices and edges in the recursion tree. In particular, the instances to the max-flow calls have $\tO(n\e^{-3}\log\De)$ vertices and edges in total.
\EP

Resetting $\e\gets\Th(\e/\log(n\De))$, we have thus proved \thm{approx-w}, restated below.
\ApproxW*

\balance

\bibliography{refs,dp-refs}

%\appendix

%\balance
%\section{Rooted minimal Gomory-Hu Steiner tree}
%\input{minimal}

\end{document}